\documentclass[aps,onecolumn,showpacs,nofootinbib,showkeys]{revtex4-2}
\usepackage[margin=2.0 cm]{geometry}

\RequirePackage[T1]{fontenc}

\usepackage{epsfig,graphicx,amsmath,amssymb,bm}
\RequirePackage{mathptmx}      % use Times fonts if available on your TeX system
\usepackage{mathrsfs}
\usepackage[english]{babel}
\usepackage{amssymb}
\RequirePackage{color}
\usepackage{changes}
\usepackage{slashed}
\usepackage{multirow}
\usepackage{scalerel}
\usepackage{tikz-feynman}
\usepackage{textcomp}
\usepackage{subcaption}
\usepackage[english]{babel}%используем русский и английский языки с переносами
\usepackage{float}
\RequirePackage{hyperref}
\hypersetup{
    linktocpage,
    colorlinks,
    citecolor=brown,
    filecolor=black,
    linkcolor=brown,
    urlcolor=purple,
}
\usepackage{nccmath}

\begin{document}

\title{Three-meson $\tau$ decays involving kaons and $\eta$ mesons in the NJL model}


\author{M.K. Volkov$^{1}$}\email{volkov@theor.jinr.ru}
\author{A.A. Pivovarov$^{1}$}\email{tex$\_$k@mail.ru}
\author{K. Nurlan$^{1,2,3}$}\email{nurlan@theor.jinr.ru}

\affiliation{$^1$ Bogoliubov Laboratory of Theoretical Physics, JINR, 
                 141980 Dubna, Russia \\
                $^2$ The Institute of Nuclear Physics, Almaty, 050032, Kazakhstan \\
                $^3$ L. N. Gumilyov Eurasian National University, Astana, 010008, Kazakhstan}   

%%%%%%%%%%%% ABSTRACT %%%%%%%%%%%%
%%%%%%%%%%%%%%%%%%%%%%%%%%%%%

\begin{abstract}
Branching fractions of decays $\tau \to K^0 \pi^- \eta\nu_\tau$, $\tau \to K^- \pi^0 \eta\nu_\tau$, $\tau \to K^-K^0 \eta\nu_\tau$ and $\tau \to K^- \eta \eta \nu_\tau$ are calculated in the $U(3)\times U(3)$ chiral NJL quark model. The contact, vector, axial-vector and pseudoscalar channels are considered. It is shown that the axial vector channel is dominant. The obtained results are in satisfactory agreement with experiment.

%\keywords{}

\end{abstract}


\maketitle

%%%%%%%%%%%%%%%%%%%%%%%%%%%%
%%%%%%%%%% INTRODUCTION %%%%%%%%%%
%%%%%%%%%%%%%%%%%%%%%%%%%%%%

\section{\label{Intro}Introduction}
The study of hadronic $\tau$ decays is of great importance for a deeper understanding of strong interactions at low energies (< 2 GeV). At this energy scale, the perturbation theory of Quantum Chromodynamics is not applicable.
Therefore, it is necessary to apply various phenomenological models. One of these models that has been successfully used for the description of low-energy meson interactions is the $U(3)\times U(3)$ chiral symmetric Nambu--Jona-Lasinio (NJL) quark model~\cite{Eguchi:1976iz,Ebert:1982pk,Volkov:1984kq,Volkov:1986zb,Ebert:1985kz,Vogl:1991qt,Klevansky:1992qe,Volkov:1993jw,Hatsuda:1994pi,Ebert:1994mf,Buballa:2003qv,Volkov:2005kw}. In the framework of this model, numerous the $\tau$ lepton decays and processes of electron-positron annihilation into meson states were successfully described~\cite{Volkov:2016umo,Volkov:2022jfr}. 

The NJL model is based on the chiral symmetry of strong interactions. This symmetry is partially broken by the current masses of the u, d, and s quarks within the limits of 15\%~\cite{Vainshtein:1970zm}. In the case of including heavier quarks, the chiral symmetry breaking becomes unacceptably strong. 
That is why, in the existing versions of the NJL model, the symmetry higher than $U(3)\times U(3)$ is not applied. In the case of $\eta$ mesons, it is necessary to take into account the mixing of light u and d quarks with a heavier s quark. 
These mixing arises when the gluon anomaly is taken into account, that is well described by using the 't Hooft interaction~\cite{tHooft:1976rip, Volkov:1998ax}. 
The processes considered here include both strange kaons and $\eta$ mesons, in the description of which the breaking of the chiral symmetry pointed out above plays a very important role. Partially, this is the reason why these processes are theoretical poorly studied at the present time.

However, experimentally, these processes have been studied better. In the work of the Belle collaboration~\cite{Belle:2008jjb}, the results of experimental measurements of partial widths of $\tau$ lepton decays containing $\eta$ mesons were obtained with a significantly higher precision than in the previous measurements.

In the present paper, we give a theoretical description of three-meson $\tau$ decays containing $K$ and $\eta$ mesons $\tau \to K^0 \pi^- \eta\nu_\tau$, $\tau \to K^- \pi^0 \eta\nu_\tau$, $\tau \to K^-K^0 \eta\nu_\tau$, $\tau \to K^- \eta \eta \nu_\tau$ in the framework of the NJL model. We take into account the contact contributions and the contributions from the intermediate axial vector, vector, and pseudoscalar states.

%%%%%%%%%%%%%%%%%%%%%%%%%%%%%%%%%%%%%%%%%%
%%%%%%%%%%%%%%%%%%%%%%%%%%%%%%%%%%%%%%%%%%
\section{Lagrangian of the NJL model}
For the calculation of the processes $\tau \to K^0 \pi^- \eta\nu_\tau$, $\tau \to K^- \pi^0 \eta\nu_\tau$, $\tau \to K^-K^0 \eta\nu_\tau$, $\tau \to K^- \eta \eta \nu_\tau$ we need the vertices containing the meson states $K$, $K^{*}$, $K_{1}$, $\pi$, $\eta$. The fragment of the quark-meson Lagrangian of the NJL model with such vertices takes the following form~\cite{Volkov:2005kw,Volkov:2022jfr}:
\begin{eqnarray}
	\Delta L_{int} & = & \bar{q}\left\{\sum_{i=0,\pm}\left[ig_{K}\gamma^{5}\lambda^{K}_{i}K^{i} + ig_{\pi}\gamma^{5}\lambda^{\pi}_{i}\pi^{i} + \frac{g_{\rho}}{2}\gamma^{\mu}\lambda^{\rho}_{i}\rho^{i}_{\mu} + \frac{g_{K^{*}}}{2}\gamma^{\mu}\lambda^{K}_{i}K^{*i}_{\mu} + \frac{g_{a_1}}{2}\gamma^{\mu}\gamma^{5}\lambda^{\rho}_{i}a^{i}_{1\mu} + \frac{g_{K_1}}{2}\gamma^{\mu}\gamma^{5}\lambda^{K}_{i}K^{i}_{1A\mu} \right]\right. \nonumber\\
	&&\left. + ig_{K}\gamma^{5}\lambda_{0}^{\bar{K}}\bar{K}^{0} + i\sin\bar{\theta}g_{\eta^{u}}\gamma^{5}\lambda^{u}\eta + i\cos\bar{\theta}g_{\eta^{s}}\gamma^{5}\lambda^{s}\eta + \frac{g_{K^{*}}}{2}\gamma^{\mu}\lambda^{\bar{K}}_{0}\bar{K}^{*0}_{\mu}\right\}q,
\end{eqnarray}
where $q$ and $\bar{q}$ are triplets of the u, d, and s quarks with the constituent masses $m_{u} \approx m_{d} = 270$~MeV, $m_{s} = 420$~MeV, $\lambda$ are the linear combinations of the Gell-Mann matrices, $\bar{\theta} = \theta^{0} - \theta$ is the mixing angle of the mesons $\eta$ and $\eta^{'}$, $\theta = -19^{\circ}$ is the deviation of the ideal mixing angle $\theta^{0} = 35.3^{\circ}$~\cite{Volkov:1998ax}.

The strange axial vector meson $K_{1A}$ appearing in the Lagrangian represents the combination of two states that are the results of the mixing of the states $K_{1A}$ and $K_{1B}$ \cite{Volkov:1984gqw, Suzuki:1993yc, Volkov:2019awd}:
\begin{eqnarray}
    K_{1A} = K_1(1270)\sin{\alpha} + K_1(1400)\cos{\alpha},
\end{eqnarray}
where $\alpha = 57^\circ$.

The quark-meson coupling constants appear from the renormalization of the Lagrangian:
\begin{eqnarray}
	&g_{\pi} = g_{\eta^{u}} = \sqrt{\frac{Z_{\pi}}{4 I_{20}}}, \quad g_{\eta^{s}} = \sqrt{\frac{Z_{s}}{4 I_{02}}}, \quad g_{K} = \sqrt{\frac{Z_{K}}{4 I_{11}}},& \nonumber\\
    &g_{\rho} = g_{a_1} = \sqrt{\frac{3}{2 I_{20}}}, \quad g_{K^{*}} = g_{K_1} = \sqrt{\frac{3}{2 I_{11}}},&
\end{eqnarray}
where
\begin{eqnarray}
    &Z_{\pi} = \left(1 - 6\frac{m_{u}^{2}}{M^{2}_{a_{1}}}\right)^{-1}, \quad Z_{s} = \left(1 - 6\frac{m_{s}^{2}}{M^{2}_{f_{1}}}\right)^{-1},& \nonumber\\
	&Z_{K} = \left(1 - \frac{3}{2}\frac{(m_{u} + m_{s})^{2}}{M^{2}_{K_{1A}}}\right)^{-1},& \nonumber\\
	&M^{2}_{K_{1A}} = \left(\frac{\sin^{2}{\alpha}}{M^{2}_{K_{1}(1270)}} - \frac{\cos^{2}{\alpha}}{M^{2}_{K_{1}(1400)}}\right)^{-1},&
\end{eqnarray}
$Z_{\pi}$, $Z_{s}$ and $Z_{K}$ are the factors describing $\pi - a_1$, $\eta - f_{1}$ and $K - K_{1}$ transitions, $M_{a_{1}} = 1230$~MeV, $M_{f_{1}} = 1426$~MeV, $M_{K_{1}(1270)} = 1253$~MeV, $M_{K_{1}(1400)} = 1403$~MeV \cite{ParticleDataGroup:2022pth} are the masses of the axial vector mesons $a_{1}$ and $K_{1}$.

The integrals in the definitions of the coupling constants and appearing in the quark loops as a result of the renormalization of the Lagrangian take the form:
\begin{equation}
\label{integral}
	I_{nm} = -i\frac{N_{c}}{(2\pi)^{4}}\int\frac{\theta(\Lambda^{2} + k^2)}{(m_{u}^{2} - k^2)^{n}(m_{s}^{2} - k^2)^{m}}
	\mathrm{d}^{4}k,
\end{equation}
where $\Lambda = 1265$~MeV is the cut-off parameter~\cite{Volkov:2022jfr}.

%%%%%%%%%%%%%%%%%%%%%%%%%%%%%%%%%%%%%%%%%%
%%%%%%%%%%%%%%%%%%%%%%%%%%%%%%%%%%%%%%%%%%
\section{The amplitudes of the decays $\tau \to K \pi \eta \nu_{\tau}$}
Since the decays under consideration are four-particle, the corresponding diagrams can contain up to two intermediate states. As the first intermediate state, the axial vector, vector, or pseudoscalar meson can be considered. Therefore, the axial vector, vector, and pseudoscalar channels can be distinguished. As the second intermediate resonance, only vector mesons may take place. Besides, contact diagrams where the W boson does not produce the first intermediate resonance but directly decays into two meson states are also possible. Since the contact diagram contains the vector and axial vector parts, the contact contributions can be included in the appropriate axial vector and vector channels. 

\begin{figure*}[t]
 \centering
  \begin{subfigure}{0.5\textwidth}
   \centering
    \begin{tikzpicture}
     \begin{feynman}
      \vertex (a) {\(\tau\)};
      \vertex [dot, right=2cm of a] (b){};
      \vertex [above right=2cm of b] (c) {\(\nu_{\tau}\)};
      \vertex [dot, below right=1.2cm of b] (d) {};
      \vertex [dot, above right=1.2cm of d] (e) {};
      \vertex [dot, below right=1.2cm of d] (h) {};
      \vertex [dot, right=1.2cm of e] (f) {};
      \vertex [dot, above right=1.0cm of f] (n) {};  
      \vertex [dot, below right=1.0cm of f] (m) {};   
      \vertex [right=1.2cm of n] (l) {\(\ K \)}; 
      \vertex [right=1.2cm of m] (s) {\(\pi (\eta) \)};  
      \vertex [right=1.4cm of h] (k) {\(\eta (\pi) \)}; 
      \diagram* {
         (a) -- [fermion] (b),
         (b) -- [fermion] (c),
         (b) -- [boson, edge label'=\(W\)] (d),
         (d) -- [fermion] (e),  
         (e) -- [fermion] (h),
         (d) -- [anti fermion] (h),
         (e) -- [edge label'=\({ K^{*}} \)] (f),
         (f) -- [fermion] (n),
         (n) -- [fermion] (m),
         (f) -- [anti fermion] (m), 
         (h) -- [] (k),
         (n) -- [] (l),
	 (m) -- [] (s),
      };
     \end{feynman}
    \end{tikzpicture}
  \end{subfigure}%
 \caption{The contact diagram of the decays $\tau \to K^* \eta (K^* \pi) \to K \pi \eta \nu_\tau$.}
 \label{diagram1}
\end{figure*}
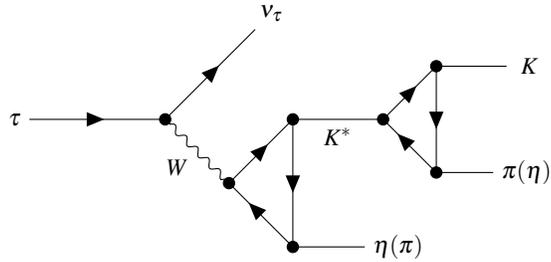

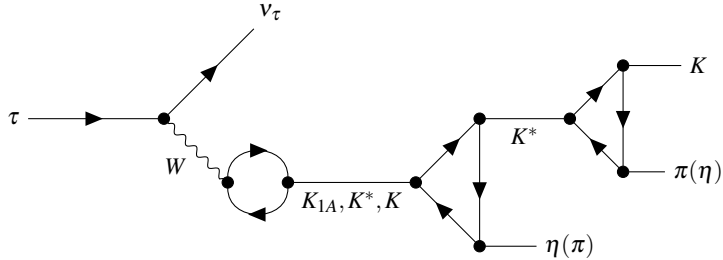
\begin{figure*}[t]
 \centering
 \centering
 \begin{subfigure}{0.5\textwidth}
  \centering
   \begin{tikzpicture}
    \begin{feynman}
      \vertex (a) {\(\tau\)};
      \vertex [dot, right=2cm of a] (b){};
      \vertex [above right=2cm of b] (c) {\(\nu_{\tau}\)};
      \vertex [dot, below right=1.2cm of b] (d) {};
      \vertex [dot, right=0.8cm of d] (l) {};
      \vertex [dot, right=1.7cm of l] (g) {};
      \vertex [dot, above right=1.2cm of g] (e) {};
      \vertex [dot, below right=1.2cm of g] (h) {};      
      \vertex [dot, right=1.2cm of e] (f) {};
      \vertex [dot, above right=1.0cm of f] (n) {};
      \vertex [dot, below right=1.0cm of f] (m) {};
      \vertex [right=1.0cm of n] (s) {\( K \)};
      \vertex [right=1.0cm of m] (r) {\( \pi (\eta) \)};
      \vertex [right=1.2cm of h] (k) {\( \eta (\pi) \)}; 
      \diagram* {
         (a) -- [fermion] (b),
         (b) -- [fermion] (c),
         (b) -- [boson, edge label'=\(W\)] (d),
         (d) -- [fermion, inner sep=1pt, half left] (l),
         (l) -- [fermion, inner sep=1pt, half left] (d),
         (l) -- [edge label'=\({ K_{1A}, K^{*}, K } \)] (g),
         (g) -- [anti fermion] (h),  
         (h) -- [anti fermion] (e),
         (e) -- [anti fermion] (g),      
         (e) -- [edge label'=\( K^{*} \)] (f),
         (f) -- [fermion] (n),
         (n) -- [fermion] (m),
         (m) -- [fermion] (f),
         (h) -- [] (k),
         (n) -- [] (s),
         (m) -- [] (r),
      };
     \end{feynman}
    \end{tikzpicture}
  \end{subfigure}%
 \caption{ The diagram with the intermediate mesons describing the decays $\tau \to K^* \eta (K^* \pi) \to K \pi \eta \nu_\tau$.}
 \label{diagram2}
\end{figure*}

The diagrams of the processes $\tau \to K \pi \eta \nu_{\tau}$ are presented in figures~\ref{diagram1} and \ref{diagram2}.

The amplitude of the process obtained in the framework of the NJL model takes the following form:
\begin{eqnarray}
\label{amplitude}
    \mathcal{M} = G_{F} V_{us} L_{\mu} \left\{\mathcal{M}_{A} + \mathcal{M}_{V} + \mathcal{M}_{P}\right\}^{\mu},
\end{eqnarray}
where $L_{\mu}$ is the weak lepton current; $\mathcal{M}_{A}$, $\mathcal{M}_{V}$ and $\mathcal{M}_{P}$ are the axial vector, vector, and pseudoscalar channels respectively, and for the process $\tau \to K^- \pi^0 \eta \nu_{\tau}$ they take the form
\begin{eqnarray}
    \mathcal{M}_{A}^{\mu} & = & -i\frac{3}{2} m_{s} Z_{K} \frac{g_{\pi}}{g_{K}} \left[\left(g^{\mu\nu}h_{K_1(1270)} - q^{\mu}q^{\nu}\right)BW_{K_1(1270)}^{q} \sin^2{\alpha} + \left(g^{\mu\nu}h_{K_1(1400)} - q^{\mu}q^{\nu}\right)BW_{K_1(1400)}^{q} \cos^2{\alpha}\right] \nonumber\\
    && \times\left\{\left[ g_{\eta^u}\sin{\bar{\theta}} +\sqrt{2} g_{\eta^s} \cos{\bar{\theta}}\right]BW_{K^{*-}}^{q_{K\eta}}\left(\frac{1}{Z_{\eta}}p_{K} - p_{\eta}\right)_{\nu} + \left[ g_{\eta^u}\sin{\bar{\theta}} +\sqrt{2} \frac{m_u}{m_s} g_{\eta^s} \cos{\bar{\theta}}\right]BW_{K^{*-}}^{q_{K\pi}}\left(\frac{1}{Z_{K_{1}}}p_{K} - \frac{1}{Z_{a_{1}}}p_{\pi}\right)_{\nu}\right\}, \nonumber\\
    \mathcal{M}_{V}^{\mu} & = & 6 m_u g_{\pi} g_K I_{cu} BW_{K^{*}}^{q} h_{K^*} \frac{1+ Z_{\eta}}{Z_{\eta}} e^{\mu \nu \lambda \delta} p_{\pi\nu} p_{K\lambda} p_{\eta\delta} \nonumber\\
    && \times\left\{\left[ g_{\eta^u}\sin{\bar{\theta}} +\sqrt{2} g_{\eta^s} \cos{\bar{\theta}}\right]BW_{K^{*-}}^{q_{K\eta}} - \left[ g_{\eta^u}\sin{\bar{\theta}} +\sqrt{2}  \frac{m_s}{m_u} \frac{I_{cs}}{I_{cu}} g_{\eta^s} \cos{\bar{\theta}}\right] BW_{K^{*-}}^{q_{K\pi}}\right\}, \nonumber\\
    \mathcal{M}_{P}^{\mu} & = & \frac{3}{4} (m_s + m_u) Z_K^2 \frac{g_{\pi}}{g_{K}}\left[ g_{\eta^u}\sin{\bar{\theta}} +\sqrt{2} g_{\eta^s} \cos{\bar{\theta}}\right] q^{\mu} BW_{K}^{q} \nonumber\\
    && \times\left\{BW_{K^{*-}}^{q_{K\eta}}\left(\frac{1}{Z_{K_{1}}}q + \frac{1}{Z_{a_{1}}}p_{\pi}\right)^{\nu}\left(\frac{1}{Z_{\eta}}p_{K} - p_{\eta}\right)_{\nu} + BW_{K^{*-}}^{q_{K\pi}}\left(\frac{1}{Z_{\eta}}q + p_{\eta}\right)^{\nu}\left(\frac{1}{Z_{K_{1}}}p_{K} - \frac{1}{Z_{a_{1}}}p_{\pi}\right)_{\nu}\right\},
\end{eqnarray}
where $p_K$, $p_{\pi}$, $p_{\eta}$ are the momenta of the final mesons, $q_{K\pi} = p_K + p_{\pi}$, $q_{K\eta} = p_K + p_{\eta}$, $q = p_K + p_{\pi} + p_{\eta}$.

The intermediate states are discribed using the Breit-Wigner propagator:
\begin{eqnarray}
    BW_{M}^{p} = \frac{1}{M_{M}^{2} - p^{2} - i\sqrt{p^{2}}\Gamma_{M}},
\end{eqnarray}
where $M$ designates a meson, $M_{M}$, $\Gamma_{M}$ and $p$ are its mass, width and momentum respectively .

The factors $Z_{K_1}$, $Z_{a_1}$ and $Z_{\eta}$ appear as a result of the explicit allowance for transitions between the axial vector and pseudoscalar states in the different diagram vertices:
\begin{eqnarray}
    &Z_{a_{1}} = \left(1 - 3\frac{m_{u}(3m_{u} - m_{s})}{M_{a_{1}}^{2}}\right)^{-1}, \quad Z_{K_{1}} = \left(1 - 3\frac{m_{s}(m_{u} + m_{s})}{M_{K_{1A}}^{2}}\right)^{-1}&, \nonumber\\
    & Z_{\eta} = \left(1 - 3\frac{m_s g_{\eta_u} \sin{\bar{\theta}} + \sqrt{2} m_u g_{\eta_s} \cos{\bar{\theta}} }{g_{\eta_u} \sin{\bar{\theta}} + \sqrt{2} g_{\eta_s} \cos{\bar{\theta}}} 
    \frac{m_s + m_u }{M^2_{K_{1A}}}\right)^{-1},&
\end{eqnarray}

The factors $h_{K_1}$ and $h_{K^*}$ appear as a result of the summation of the diagrams with intermediate mesons and the appropriate parts of the contact diagrams:
\begin{eqnarray}
    h_{K_1} & = & M^2_{K_1} - i\sqrt{q^{2}}\Gamma_{K_1} - \frac{3}{2} \left(m_s + m_u\right)^2, \nonumber\\
    h_{K^*} & = & M^2_{K^*} - i\sqrt{q^{2}}\Gamma_{K^*} -\frac{3}{2} \left(m_s - m_u\right)^2.
\end{eqnarray}

In the vector channel, the combinations of the convergent integrals appear:
\begin{eqnarray}
    I_{cu} & = & I_{21} + m_{u}(m_{s} - m_{u})I_{31}, \nonumber\\
    I_{cs} & = & I_{12} - m_{s}(m_{s} - m_{u})I_{13},
\end{eqnarray}
where $I_{21}$, $I_{31}$, $I_{12}$ and $I_{13}$ were defined in (\ref{integral}).

The partial width of this decay calculated by using the above amplitude takes on the following value:
\begin{eqnarray}
    Br(\tau \to K^- \pi^0 \eta \nu_{\tau}) = (3.9 \pm 0.6) \times 10^{-5}
\end{eqnarray}

This result does not contradict the experimental value within the errors~\cite{ParticleDataGroup:2022pth}:
\begin{eqnarray}
    Br(\tau \to K^- \pi^0 \eta \nu_{\tau})_{exp} = (4.8 \pm 1.2) \times 10^{-5}
\end{eqnarray}

The amplitude of the process $\tau \to \bar{K}^0 \pi^- \eta \nu_{\tau}$ almost coincides with the amplitude of the process $\tau \to K^- \pi^0 \eta \nu_{\tau}$. The only difference is the additional factor $\sqrt{2}$. Besides, in one of the subprocesses, the neutral meson $K^{*0}$ takes place as the second intermediate resonance. As a result, one can obtain the following value for the partial decay width of this process:
\begin{eqnarray}
    Br(\tau \to \bar{K}^0 \pi^- \eta \nu_{\tau}) = (7.8 \pm 1.2) \times 10^{-5}
\end{eqnarray}

This result is also consistent with the experimental data within the theoretical and experimental uncertainties~\cite{ParticleDataGroup:2022pth}:
\begin{eqnarray}
    Br(\tau \to \bar{K}^0 \pi^- \eta \nu_{\tau})_{exp} = (9.4 \pm 1.5) \times 10^{-5}
\end{eqnarray}

%%%%%%%%%%%%%%%%%%%%%%%%%%%%%%%%%%%%%%%%%%
%%%%%%%%%%%%%%%%%%%%%%%%%%%%%%%%%%%%%%%%%%
\section{The amplitudes of decays $\tau \to K [K \eta, \eta\eta]\nu_{\tau}$} 
The amplitudes of $\tau \to K [K \eta, \eta\eta]\nu_{\tau}$ decays, as in the case of the $\tau \to K \pi \eta \nu_{\tau}$ decay, contain contributions from contact diagrams and diagrams with intermediate axial-vector, vector, and pseudoscalar mesons. The decay of $\tau \to K K \eta \nu_{\tau}$, unlike other processes, proceeds through intermediate non-strange states $a_1$, $\rho$ and $\pi$. In this process, the second resonances are $K^{*-}$ and $K^{*0}$ mesons. In the case of decay with the production of two $\eta$ mesons $\tau \to K \eta\eta\nu_{\tau}$, intermediate channels with strange mesons operate, in which the second resonance is only the $K^{*-}$ meson.

The process amplitude of $\tau \to K^- K^0 \eta \nu_{\tau}$ can be represented as (\ref{amplitude}). Then the corresponding contributions take the form
\begin{eqnarray}
\label{contributions_KKeta}
    \mathcal{M}_{A}^{\mu} & = & i\frac{3\sqrt{2}}{4}(3m_u -m_s) Z_{K} \frac{V_{ud}}{V_{us}} \left[ g_{\eta_u}\sin{\bar{\theta}} +\sqrt{2} g_{\eta_s}\cos{\bar{\theta}}\right] BW_{a_1}^{q} \left[g^{\mu\nu}h_{a_1} - q^{\mu}q^{\nu}\right] 
    \nonumber\\    
    && \times\left\{
    BW_{K^{*-}}^{q_{K^{-}\eta}}\left(\frac{1}{Z_{\eta}} p_{K^-} - p_{\eta} \right)_{\nu} +
    BW_{K^{*0}}^{q_{K^{0}\eta}}\left(\frac{1}{Z_{\eta}} p_{K^0} - p_{\eta} \right)_{\nu}    
    \right\}, \nonumber\\
    \mathcal{M}_{V}^{\mu} & = & \sqrt{2} Z_K g^2_{K^*} \frac{V_{ud}}{V_{us}} m_u I_{cu} h_{\rho} \left[ g_{\eta_u}\sin{\bar{\theta}} +\sqrt{2} g_{\eta_s}\cos{\bar{\theta}}\right] \frac{1+ Z_{\eta_1}}{Z_{\eta_1}} BW_{\rho}^q 
   \left\{ BW_{K^{*-}}^{q_{K^{-}\eta}} + BW_{K^{*0}}^{q_{K^{0}\eta}} \right\} 
    \nonumber\\
    && \times e^{\mu \nu \lambda \delta} p_{K^-\nu} p_{K^{0}\lambda} p_{\eta\delta}, \nonumber\\
    \mathcal{M}_{P}^{\mu} & = & i \frac{3\sqrt{2}}{2} m_u \frac{V_{ud}}{V_{us}} Z_K \left[ g_{\eta_u}\sin{\bar{\theta}} +\sqrt{2} g_{\eta_s}\cos{\bar{\theta}}\right] q^\mu BW_{\pi}^q \nonumber \\
     && \times\left\{
    BW_{K^{*-}}^{q_{K^{-}\eta}} {\left( \frac{1}{Z_{K_1}}p_{K^0} +  \frac{1}{Z_{a_1}}q \right)}_{\nu} \left(\frac{1}{Z_{\eta}} p_{K^-} - p_{\eta} \right)_{\nu} +
    BW_{K^{*0}}^{q_{K^{0}\eta}} {\left( \frac{1}{Z_{K_1}}p_{K^-} + \frac{1}{Z_{a_1}}q \right)}_{\nu} \left(\frac{1}{Z_{\eta}} p_{K^0} - p_{\eta} \right)_{\nu},    
    \right\}
\end{eqnarray}
where $p_{K^{0}}$, $p_{K^{-}}$ and $p_{\eta}$ are meson momenta in the final states, $q = p_{K^{0}} + p_{K^{-}} + p_{\eta}$ is the momentum of the first intermediate meson, $q_{K^{-}\eta} = p_{K^{-}} + p_{\eta}$, $q_{K^{0}\eta} = p_{K^{0}} + p_{\eta}$. The factors $h_{a_1}$ and $h_{\rho}$ take the form
\begin{eqnarray}
    h_{a_1} & = & M^2_{a_1} - i\sqrt{q^{2}}\Gamma_{a_1} - 6 m^2_u, \nonumber\\
    h_{\rho} & = & M^2_{\rho} - i\sqrt{q^{2}}\Gamma_{\rho}.
\end{eqnarray}

Using the obtained amplitude, we get the following estimate for the branching fraction of $\tau \to K K \eta \nu_{\tau}$

\begin{eqnarray}
    Br(\tau \to K K^- \eta \nu_{\tau}) = (1.6 \pm 0.2) \times 10^{-6}.
\end{eqnarray}

This result does not exceed the experimental limit for the branching fraction ~\cite{ParticleDataGroup:2022pth}:

\begin{eqnarray}
    Br(\tau \to K K^- \eta \nu_{\tau})_{exp} < 9.0 \times 10^{-6}.
\end{eqnarray}

The axial-vector, vector, and pseudoscalar channels give the following contributions to the decay amplitude of $\tau \to K^* \eta \to K\eta\eta\nu_\tau$

\begin{eqnarray}
\label{contributions_Ketaeta}
    \mathcal{M}_{A}^{\mu} & = &- i \frac{3}{2}\frac{Z_K}{g_K} m_s \left( g_{\eta^u}\sin{\bar{\theta}} +\sqrt{2} \frac{m_u}{m_s} g_{\eta^s} \cos{\bar{\theta}}\right) \left( g_{\eta_u} \sin{\bar{\theta}} + \sqrt{2} g_{\eta_s} \cos{\bar{\theta}}\right)
        \nonumber\\   
    && \times \left[ \left( g^{\mu\nu} h_{K_1(1270)} - q^\mu q^\nu \right)BW_{K_1(1270)}^{q} \sin^2{\alpha} + \left( g^{\mu\nu} h_{K_1(1400)} - q^\mu q^\nu \right) BW_{K_1(1400)}^{q} \cos^2{\alpha}\right]   
    \nonumber\\   
    && \times BW^{qK\eta}_{K^{*}} \left(\frac{1}{Z_{\eta}} p_{K} - p^{(1)}_{\eta} \right)_{\nu}
    + \left( p^{(1)}_{\eta} \leftrightarrow p^{(2)}_{\eta} \right)   
    , \nonumber\\
    \mathcal{M}_{V}^{\mu} & = & 6 g_K m_u I_{cu} \left( g_{\eta^u}\sin{\bar{\theta}} +\sqrt{2}  \frac{m_s}{m_u} \frac{I_{cs}}{I_{cu}} g_{\eta^s} \cos{\bar{\theta}}\right) \left( g_{\eta_u} \sin{\bar{\theta}} + \sqrt{2} g_{\eta_s} \cos{\bar{\theta}}\right) \frac{Z_{\eta} +1}{Z_{\eta}}
BW^q_{K^{*}} h_{K^*} BW^{qK\eta}_{K^{*}} 
      \nonumber\\   
    && \times e^{\mu \nu \lambda \delta} p^{(1)}_{\eta \nu} p_{K^-\lambda} p^{(2)}_{\eta\delta} + \left( p^{(1)}_{\eta} \leftrightarrow p^{(2)}_{\eta} \right)
  \nonumber\\
    \mathcal{M}_{P}^{\mu} & = & \frac{3}{4} (m_s + m_u) \frac{Z^2_K}{g_K} {\left( g_{\eta_u} \sin{\bar{\theta}} + \sqrt{2} g_{\eta_s} \cos{\bar{\theta}}\right)}^2  q^\mu BW_K^{q} BW^{qK\eta}_{K^{*}} {\left( \frac{1}{Z_\eta}q + p^{(1)}_\eta \right)}_\nu {\left( \frac{1}{Z_\eta}p_K - p^{(2)}_\eta \right)}_\nu +\left( p^{(1)}_{\eta} \leftrightarrow p^{(2)}_{\eta} \right)
\end{eqnarray}  

The obtained estimate for the branching fraction of $\tau \to K\eta\eta\nu_\tau$ in the NJL model is
\begin{eqnarray}
    Br(\tau \to K \eta \eta \nu_{\tau}) = 1.0 \times 10^{-8}.
\end{eqnarray}

The model prediction for the $\tau \to K\eta\eta\nu_\tau$ decay also does not exceed the experimental limit for the branching fraction~\cite{ParticleDataGroup:2022pth}:
\begin{eqnarray}
    Br(\tau \to K \eta \eta \nu_{\tau})_{exp} < 3.0 \times 10^{-6}.
\end{eqnarray}

%%%%%%%%%%%%%%%%%%%%%%%%%%%%%%%%%%%%%%%%%%
%%%%%%%%%%%%%%%%%%%%%%%%%%%%%%%%%%%%%%%%%%
\section{Conclusion}
In this paper, within the standard $U(3) \times U(3)$ chiral quark NJL model, a theoretical descriptions of $\tau$ lepton decays into three pseudoscalar mesons involving a kaon and an $\eta$ meson in the final state are given. The contributions from contact channels and intermediate channels with axial-vector, vector, and pseudoscalar mesons are considered. The calculations show that the axial-vector channels play a decisive role in all the cases considered. The mixing of the $K_{1A}$ and $K_{1B}$ states is taken into account in the axial-vector channel with $K_1(1270)$ and $K_1(1400)$ intermediate mesons. The obtained results are in satisfactory agreement with the experimental data within the experimental and theoretical uncertainties.

As regards the estimate of the contribution from the box diagrams, taking it into account in the pseudoscalar decay channel of the decay $\tau \to K^- \pi^0 \eta \nu_\tau$ gives the value $Br(\tau \to K^- \pi^0 \eta \nu_\tau)_{box} = 6.4 \times 10^{-8}$, which is three orders of magnitude lower than the experimental result. The box diagram in the vector channel does not exceed the contribution of the rest of the vector channels, which itself, as a rule, is two orders of magnitude lower than the experiment. 

They were not explicitly taken into account in the amplitude due to their small contributions. Taking into account the box diagram in the axial-vector channel leads to going beyond the framework of the NJL model, that was formulated in the lowest order in terms of the quark-meson coupling constants. This order corresponds to logarithmic divergent terms at the diagram vertices. The exception is anomalous vertices, which should be taken into account in this approximation. In the axial-vector channel, the box diagram does not contain divergent integrals and is not an anomalous vertex. Therefore, it goes beyond the NJL model approximation considered here.

From a theoretical point of view, the decay of $\tau \to K^- \pi^0 \eta \nu_\tau$ was previously described in \cite{Pich:1987qq}. However, a relatively small branching fraction $Br(\tau \to K^- \pi^0 \eta \nu_\tau)=8.8 \times 10^{-6}$ was obtained there. This was a consequence of the assumption that the vector channel gives the main contribution. This resulted in a small value for the partial decay width. Among other works close to the description of the decays considered here, we can note the paper \cite{Li:1996md}. 
In this work, the decay of $\tau \to K^* \eta \nu_\tau$ was described, which in our case is an intermediate process for the decays $\tau \to K \pi \eta \nu_\tau$ and $\tau \to K^- \eta \eta \nu_\tau$. It was determined there that the axial-vector channel is dominant, and the branching fraction $Br(\tau \to K^{*-} \eta \nu_\tau)=1.01 \times 10^{-4}$ was obtained. 
Calculations in the NJL model for this decay also showed the decisive role of the axial-vector channel with the branching fraction $Br(\tau \to K^{*-} \eta \nu_\tau) = (1.23 \pm 0.18) \times 10^{-4}$~\cite{Volkov:2019izp} at experimental value $Br(\tau \to K^{*-} \eta \nu_\tau)=(1.38 \pm 0.15) \times 10^{-4}$~\cite{ParticleDataGroup:2022pth}. Thus, one of the main results of our work is the confirmation of the dominant role of the axial-vector channel in these processes.

%%%%%%%%%%%%%%%%%%%%%%%%%%%%%%%%%%%%%%%%%%
\subsection*{Acknowledgments}
This research has been funded by the Science Committee of the Ministry of Science and Higher Education of the Republic of Kazakhstan (Grant No. AP15473301).

%%%%%%%%%%%%%%%%%%%%%%%%%%%%%
%%%%%%%%%%% BIBLIOGRAPHY %%%%%%%%%%%
%%%%%%%%%%%%%%%%%%%%%%%%%%%%%

\end{document}